\documentclass[letter]{aa}  
\usepackage{graphicx}
\usepackage{txfonts}
\usepackage[bookmarksopen=false,bookmarksnumbered=true,breaklinks=true,colorlinks=true,linkcolor=blue,citecolor=blue]{hyperref} 
\usepackage{natbib}
\begin{document} 
\newcommand{\postref}[1]{{#1}}
\newcommand{\postrefref}[1]{{#1}}

   \title{Origin of the asymmetry of the wind driven halo observed in high contrast images}

   \author{F.~Cantalloube\inst{1} 
		\and E.H.~Por\inst{2} 
        \and K.~Dohlen\inst{3} 
        \and J.-F.~Sauvage\inst{3,4}
        \and A.~Vigan\inst{3}
        \and M.~Kasper\inst{5}    
       	\and N.~Bharmal\inst{6} 
        \and Th.~Henning\inst{1}
        \and W.~Brandner\inst{1}
        \and J.~Milli\inst{7}
        \and C.~Correia\inst{3}
        \and T.~Fusco\inst{3,4}
        }
	  
   \institute{Max Planck Institute for Astronomy, K\"onigstuhl 17, D-69117 Heidelberg, Germany
	   		\email{cantalloube@mpia.de}    
	     	\and Leiden Observatory, Leiden University, PO Box 9513, 2300 RA Leiden, The Netherlands
	    	\and Aix Marseille Univ, CNRS, CNES, LAM, Marseille, France
            \and ONERA – The French Aerospace Lab, 92322 Châtillon, France
	 		\and European Southern Observatory (ESO), Karl-Schwarzschild-Str. 2, 85748 Garching, Germany
	 		\and Centre for Advanced Instrumentation, Durham University, South Road, Durham DH1 3LE, United Kingdom
			\and European Southern Observatory (ESO), Alonso de C\'ordova 3107, Vitacura, Casilla 19001, Santiago, Chile
        	   }

   \date{Received ; accepted }
 
  \abstract
   {The latest generation of high contrast instruments dedicated to exoplanets and circumstellar disks imaging are equipped with extreme adaptive optics and coronagraphs to reach contrasts of up to $10^{-4}$ at a few tenths of arc-seconds in the near infrared. The resulting image shows faint features, only revealed with this combination, such as the wind driven halo. The wind driven halo is due to the lag between the adaptive optics correction and the turbulence speed over the telescope pupil. However we observe an asymmetry of this wind driven halo that was not expected when the instrument was designed. In this letter, we describe and demonstrate the physical origin of this asymmetry and support our explanation by simulating the asymmetry with an end-to-end approach. \postrefref{From this work, we found out that the observed asymmetry is explained by the interference between the AO-lag error and scintillation effects, mainly originating from the fast jet stream layer located at about 12 km in altitude.} \postref{Now identified and interpreted, this effect can be taken into account for further design of high contrast imaging simulators, next generation or upgrade of high contrast instruments, predictive control algorithms for adaptive optics or image post-processing techniques.}}

   \keywords{Instrumentation: adaptive optics, Instrumentation: high angular resolution, Atmospheric effects, Techniques: image processing, Methods: data analysis, Infrared: planetary systems}

   \maketitle
%

\section{Introduction}
With the wake of the new generation of high contrast imaging (HCI) instruments equipped with extreme adaptive optics (XAO) and advanced coronagraphs, dedicated to exoplanet and circumstellar disk imaging, we can now visualize optical effects that were expected but never revealed before. 
On 8-m class telescopes, instruments such as VLT/SPHERE \citep{Beuzit2008}, Gemini/GPI \citep{Macintosh2008gpi}, Clay/MagAO-X \citep{Close2012SPIEMagAO, Males2014MagAO} and Subaru/\postref{SCExAO} \citep{Jovanovic2015scexao} are equipped with XAO, providing a Strehl ratio up to 95\% in the near infrared, and coronagraphs, providing a raw contrast of up to $10^{-4}$ at a few hundred milliarcseconds (mas). Images obtained with these instruments show features such as the correction radius of the XAO, the deformable mirror actuator grid print-through, the bright central spot due to diffraction effects in the Lyot coronagraph (Poisson spot or Arago spot), and the wind driven halo due the temporal lag between the application of the XAO correction and the evolving turbulence. All these features were expected and taken into account when designing and simulating the instrument. 

\postref{However, some unexpected features are also visible within HCI images: the wind driven halo often shows an asymmetry, one wing being brighter and broader than the other, and the point-spread function (PSF) sometimes breaks up, leading to catastrophic loss of performance. While the latter, known as the low wind effect, is described elsewhere \citep{Milli2018lwe}, describing and understanding the asymmetric wind driven halo, which also limits the high contrast capabilities of the instrument, is the object of this letter.}

In this letter, we first describe qualitatively the observed asymmetry of the wind driven halo (Sect.~\ref{sec-obs}). Based on these observations we propose an explanation and derive its mathematical demonstration (Sect.~\ref{sec-math}). To prove our interpretation, we perform end-to-end simulations taking into account the optical effect that generates the asymmetry and checked that the asymmetry indeed varies as expected with the parameters upon which it depends (Sect.~\ref{sec-simu}).

\section{Description of the observed asymmetry}
\label{sec-obs}
The wind driven halo (WDH) is the focal plane expression of the AO servolag error (also often referred to as temporal bandwidth error). The AO-lag temporal error appears when the turbulence equivalent velocity above the telescope pupil (defined via the coherence time $\tau_0$, up to a few tens of milliseconds under good conditions) is faster than the adaptive optics correction loop frequency \citep[being about $1.4~\mathrm{kHz}$ for SAXO, the XAO of SPHERE,][]{Petit2014}. Using a coronagraph and a sufficiently long detector integration time (DIT) reveals, in the focal plane, the starlight \postref{diffracted} by this specific error. As a consequence, the PSF is elongated along the projected wind direction, making a \emph{butterfly} shaped halo appear on the images. By definition, this aberration being a phase shift in the pupil plane, it must be symmetric in the focal plane. In practice however, we observe an asymmetry of the WDH along its axis: one \emph{wing} being smaller and fainter than the other.

\begin{figure*}
\resizebox{\hsize}{!}{\includegraphics{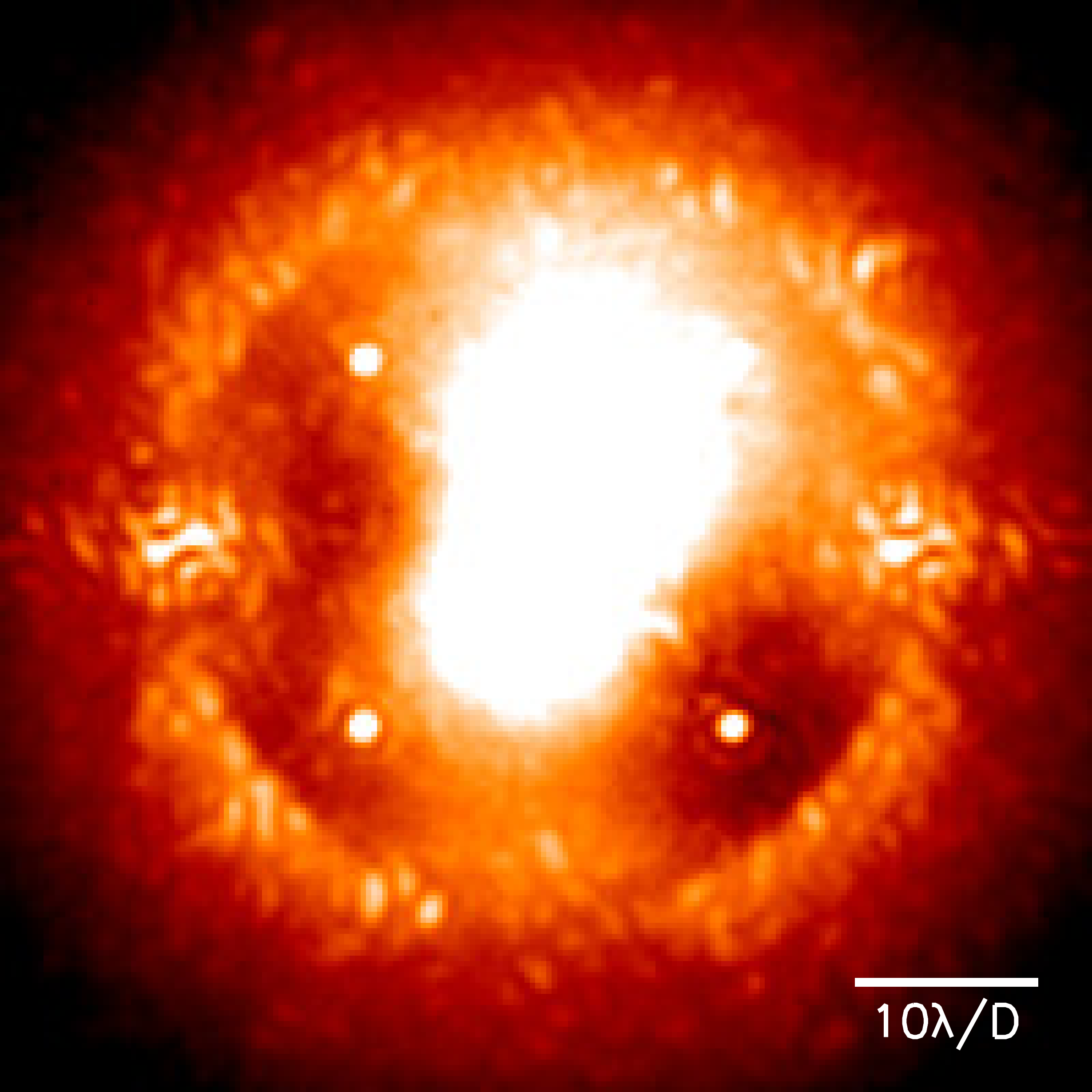}
\includegraphics{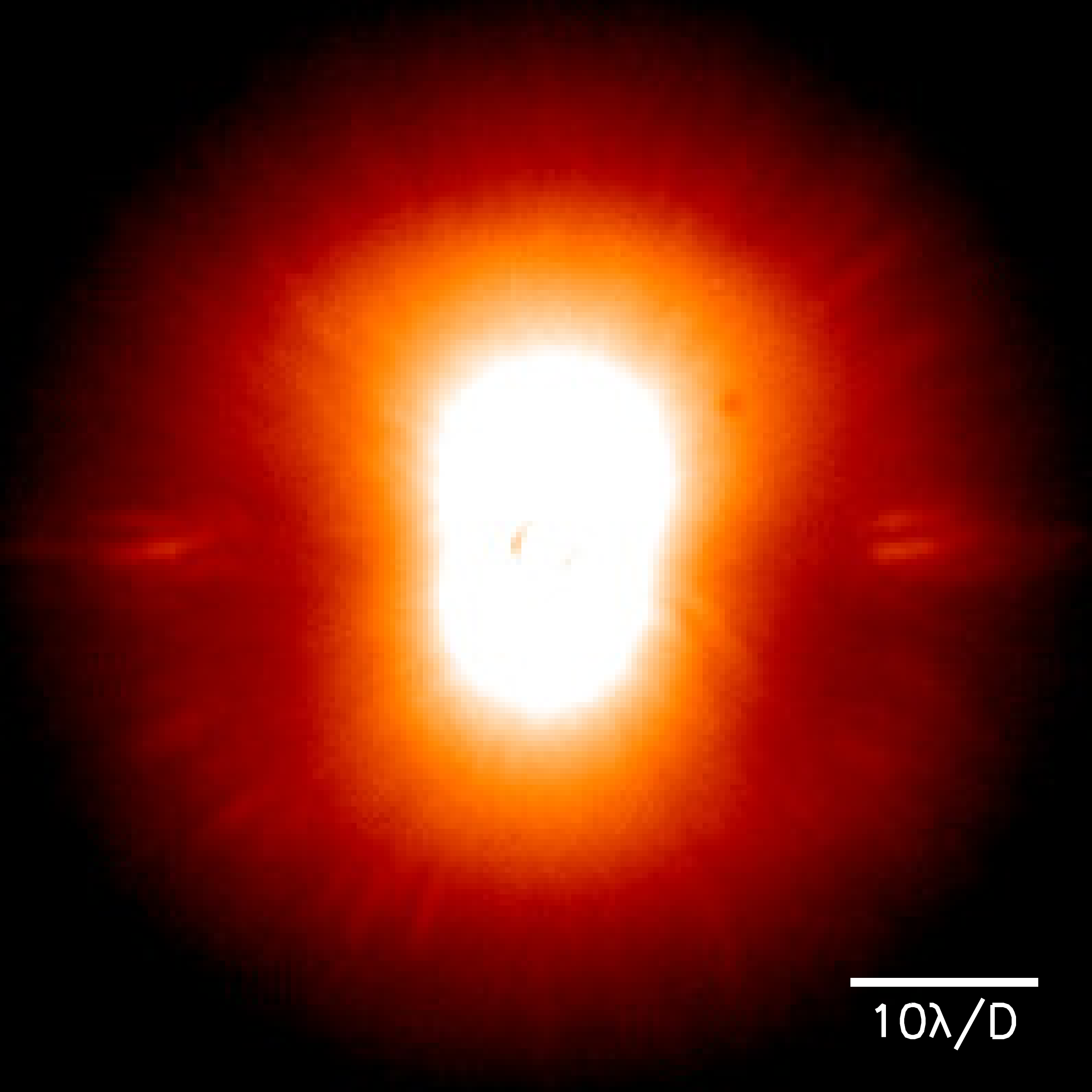}
\includegraphics{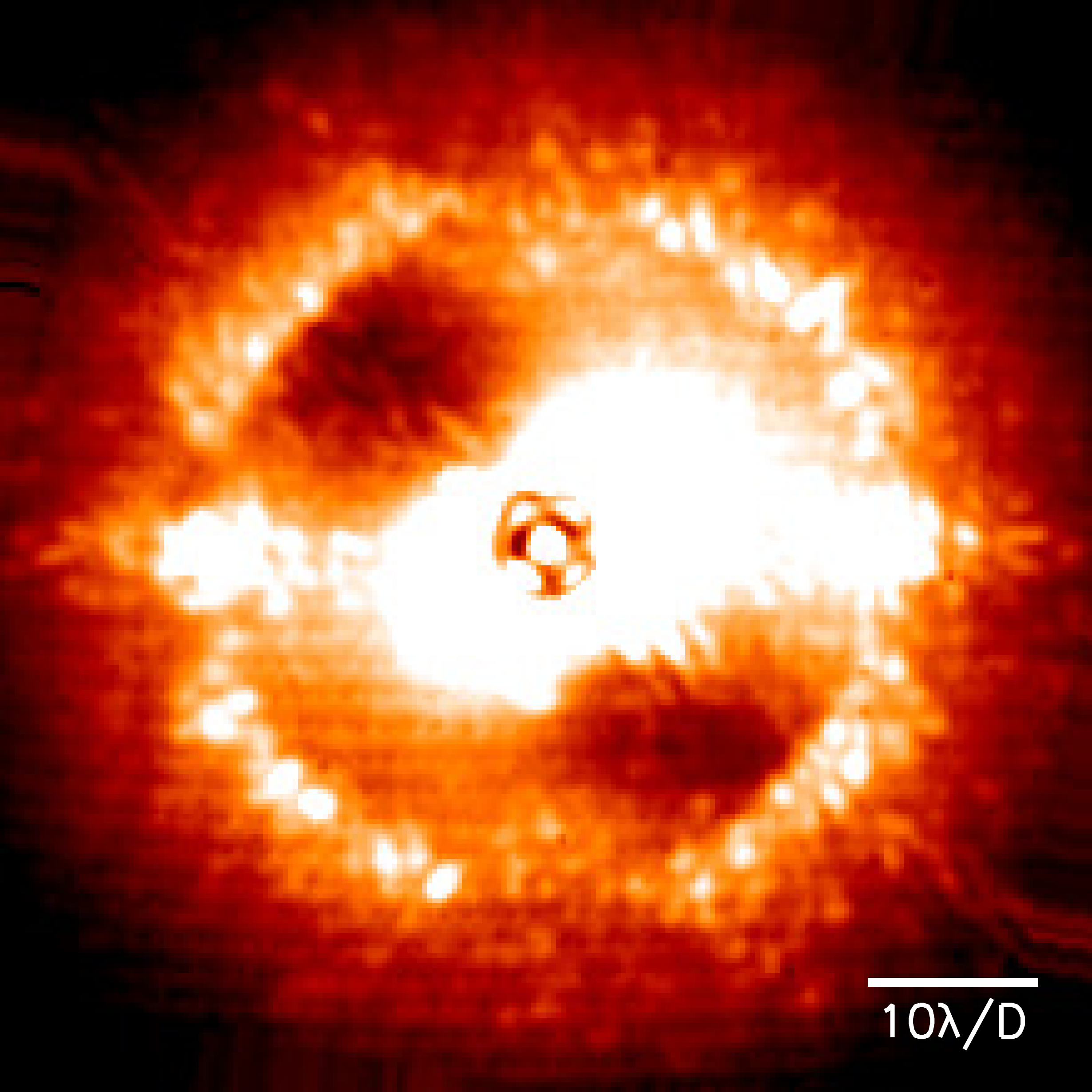}
\includegraphics{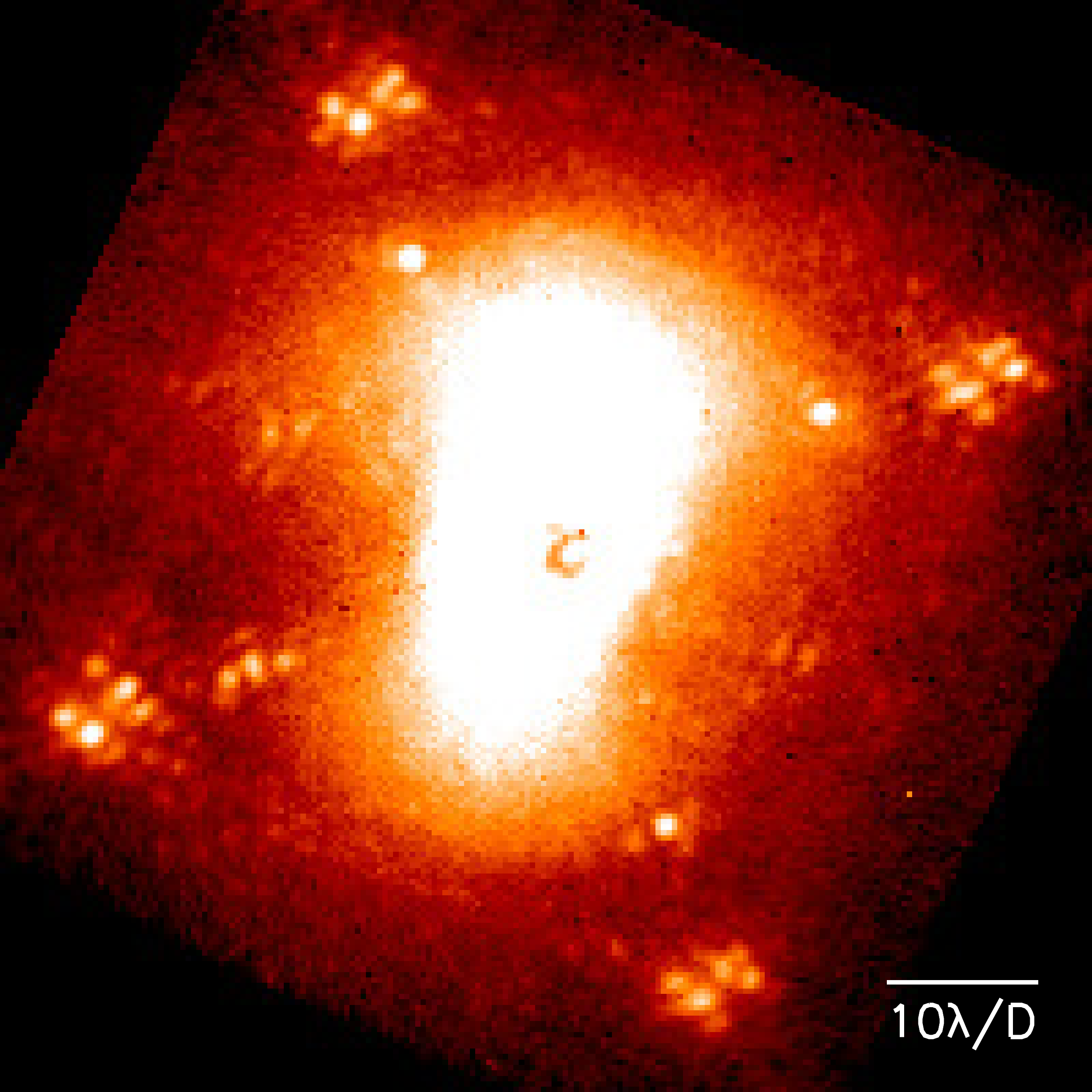}}
\caption{Coronagraphic focal plane images showing the asymmetry of the wind driven halo. 
Left: One exposure obtained with SPHERE-IRDIS (H2 band, $1.593~\mathrm{\mu m}$, $\Delta \lambda \approx 53~\mathrm{nm}$). 
Middle-Left: One exposure obtained with SPHERE-IFS (second channel of YH mode, $0.991~\mathrm{\mu m}$, $\Delta \lambda \approx 30~\mathrm{nm}$).
Middle-Right: One exposure obtained with SPHERE-IRDIS in broadband (H band, $1.625~\mathrm{\mu m}$, $\Delta \lambda \approx 291~\mathrm{nm}$).
Right: One exposure obtained with GPI (second channel of YH mode, $1.503~\mathrm{\mu m}$, $\Delta \lambda \approx 45~\mathrm{nm}$). 
The images are purposely stretched in intensity to highlight the asymmetry (log scale).}
\label{fig-AH}
\end{figure*}

\begin{figure}
\resizebox{\hsize}{!}{\includegraphics{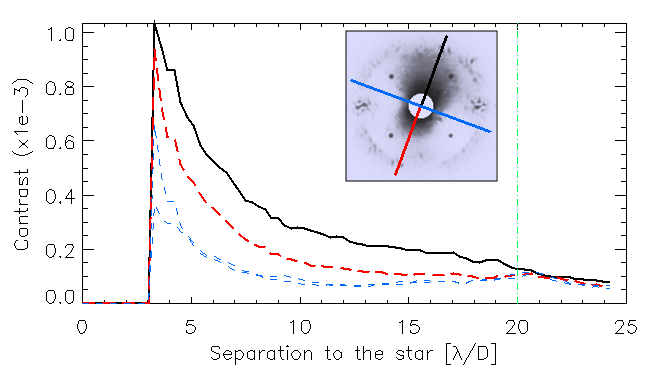}}
\resizebox{\hsize}{!}{\includegraphics{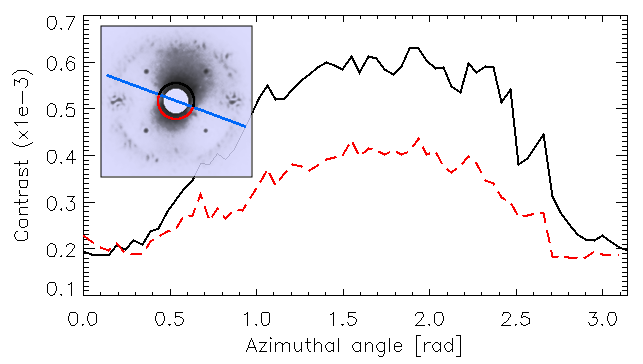}}
\caption{Profiles of the wind driven halo showing the asymmetry in a SPHERE-IRDIS image: solid line is along the brighter and bigger wing, dashed line is along the fainter and smaller wing. 
Top: Radial profile along the WDH direction (black solid and red long-dashed lines) and its perpendicular direction (blue dashed lines). The DM cutoff frequency is at $20~\mathrm{\lambda/D}$ (green dot-dashed line). 
Bottom: Azimuthal profile at $6~\mathrm{\lambda/D}$ from the star.}
\label{fig-prof}
\end{figure}

Fig.~\ref{fig-AH} shows images obtained with SPHERE and GPI\postref{\footnote{SPHERE images have been published in respectively \cite{Bonnefoy2018,Wahhaj2015,Samland2017} and GPI data in \cite{Rameau2016}.}} in which we can visualize the asymmetry of the WDH. To highlight the asymmetry, Fig.~\ref{fig-prof} shows the radial profile along the wind direction and the azimuthal profile at $6~\mathrm{\lambda/D}$ of the SPHERE-IRDIS image presented on Fig.~\ref{fig-AH} (left). 

By definition, the WDH is produced by high wind speed turbulent layers. It is now confirmed \postrefref{that} it is mainly triggered by the high altitude jet stream layer, located in a narrow region of the upper troposphere, at about $12~\mathrm{km}$ altitude above sea level ($200~\mathrm{mbar}$) which has a wind speed from $20~\mathrm{m/s}$ up to $50~\mathrm{m/s}$ \citep{Tokovinin2003,Osborn2018}. \cite{Madurowicz2018SPIE} demonstrated it by correlating the WDH direction with the wind direction at different altitude given by turbulence profiling data for the whole GPIES survey data \citep{Macintosh2013}. A forthcoming paper will similarly analyze the WDH within SPHERE data and come to the same conclusion. 

Focal plane asymmetries can only be created by combining phase and amplitude aberrations. 
As we observed that the asymmetry is pinned to the servolag signature (butterfly shape), we considered it may be caused by the interaction between servolag errors and amplitude errors created by scintillation, where the phase errors generated by high atmospheric layers propagate into amplitude errors following Fresnel's propagation laws. 

\section{Interference between scintillation and temporal error}
\label{sec-math}
In the following we provide an analytical demonstration that the combination of two well-known effects, the AO loop delay (servolag error) and scintillation (amplitude error), will indeed create the asymmetric starlight distribution observed in the high contrast images. 

In the pupil plane, the electric field can be written as:
\begin{equation}
E = (1-\epsilon) . e^{i\phi},
\end{equation}
where $\epsilon$ is the amplitude aberration and $\phi$ the phase aberration. 
An adaptive optic system measures the phase $\phi(t)$ at a given time $t$ via the wavefront sensor (WFS) and corrects it using a deformable mirror (DM). However, between the analysis of the WFS information taken at an instant $t$ and the command sent to the DM at an instant $t+\Delta t$, if the incoming turbulent phase has varied during $\Delta t$, a temporal phase error will remain (the AO servolag error). \postrefref{As a general rule, this absolute time delay $\Delta t$ varies with both the AO-loop gain and the AO-loop speed and is intrinsic to any AO system.} 
The remaining phase error $\Delta \phi$ can be written as a function of this absolute time delay $\Delta t$ following (in a closed loop system):
\begin{equation}
\Delta \phi = \phi(t) - \phi(t-\Delta t) \sim \Delta t \, \phi',
\end{equation}
where $\phi '$ is the time derivative of the phase. This approximation is valid for spatial frequencies affected by the servolag error, that is to say much lower than $1 / (v_{wind}.\Delta t)$ under the frozen flow hypothesis (i.~e. only the wind speed is responsible for the turbulent phase variation). 

Thus, after the AO correction, the electric field becomes:
\begin{equation}
\Delta E = (1- \epsilon) . e^{i (\Delta t \, \phi’)}.
\end{equation}
Which, under the small phase and small amplitude errors approximation, simplifies to:
\begin{equation}
\Delta E \simeq (1- \epsilon) . (1 + i (\Delta t . \phi')) \sim  1 - \epsilon + i \, (\Delta t \, \phi’).
\end{equation}

Seen through a perfect coronagraph (the patterns exclusively due to diffraction effects of a plane wavefront by the entrance pupil are entirely removed), the post-AO electric field $\Delta E_c$ is transformed into:
\begin{equation}
\Delta E_c \sim -\epsilon +  i \, \Delta t \, \phi’.
\label{eq-ec}
\end{equation}

The Earth’s turbulent atmosphere is present to different degrees throughout the three dimensions of the atmosphere. Fresnel propagation translates phase variations in the upper atmosphere into amplitude variations via the Talbot effect, creating the so-called scintillation. 
By the formalism of \cite{zhou2010analysis}, the phase variations in an atmospheric layer located at altitude $z$ produces an amplitude distribution at the telescope pupil of:
\begin{equation}
\epsilon = sin(2\pi \frac{z}{z_T}) \, \phi 
\end{equation}
where $z_T$ is the Talbot length, defined as the distance from which the phase error is fully converted into amplitude error: $z_T \, \dot{=} \, 1/(2f^2 \lambda)$ where $f$ is the spatial frequency and $\lambda$ the wavelength. For SPHERE, the high\postref{est} imaging wavelength is $2.2~\mathrm{\mu m}$ (K-band) and the highest corrected spatial frequency is $2.5m^{-1}$, given by the DM inter-actuator spacing ($40\times40$ actuators over the $8~\mathrm{m}$ diameter telescope pupil), yielding a minimum Talbot length of about $36~\mathrm{km}$ altitude, which is above the highest turbulence layers. This explains why, for both GPI and SPHERE, this effect \postref{was} neglected when designing the instrument.

Adding the scintillation into the coronagraphic post-AO electric field of Eq.~(\ref{eq-ec}) gives:
\begin{equation}
\Delta E_c \sim - sin(2\pi \frac{z}{z_T}) \, \phi + i \, \Delta t \, \phi’.
\end{equation}

The resulting intensity observed at the focal plane, $I_c$, is, within the Fraunhofer framework, the squared modulus of the Fourier transform of the electric field $\Delta E_c$:
\begin{equation}
I_c = | FT[ \Delta E_c ] |^2 = |  - sin(2\pi \frac{z}{z_T}) \, FT[\phi] + i \Delta t \, FT[\phi '] |^2
\label{eq-Ic}
\end{equation}

With $FT[\phi ’] = \frac{\partial FT[\phi]}{\partial t} = FT '[\phi]$ being the time derivative of the Fourier transform of the phase\postrefref{. If} we assume an arbitrary phase whose general expression can be written $\phi = \exp(i 2\pi f.r)$, $f$ being the spatial frequency and $r$ the position, then by making the change of variable \postrefref{$r \leftarrow r+\Delta r$ where we define the beam shift factor $\Delta r = (v_{wind} \, .\,\Delta t$)} to account for the servolag shift (under the frozen flow hypothesis), \postref{Eq.~(\ref{eq-Ic}) becomes}:
\begin{equation}
\begin{split}
I_c & = |  - sin(2\pi \frac{z}{z_T}) \, FT[\phi] - 2\pi \, f \, v_{wind} \, \Delta t \, FT[\phi]  |^2 \\
& = |FT[\phi]|^2 \left(sin(2\pi \frac{z}{z_T}) +  2\pi \, f \, \Delta r\right)^2,
\end{split}
\label{eq-fin}
\end{equation}
where $ |FT[\phi]|^2$ is by definition the power spectral density of the turbulent phase \postrefref{and $\Delta r$ represents the physical spatial shift between the turbulent layer and the AO correction.}
\postref{Developing Eq.~(\ref{eq-fin}) leads to} an asymmetric function of the spatial frequency $f$: $I_c$ indeed shows an asymmetric distribution of light in the high contrast images \postref{with respect to the center, originating from interferences. 
Therefore, the intensity of each wing of the WDH can be written}, for constructive and destructive interference $I_+$ and $I_-$ respectively:
\begin{align}
I_+ &= |FT[\phi]|^2 \left(sin(2\pi\frac{z}{z_T}) + 2\pi \, f \, \Delta r \right)^2; \\
I_- &= |FT[\phi]|^2 \left(sin(2\pi\frac{z}{z_T}) - 2\pi \, f \, \Delta r \right)^2.
\end{align}
We thus demonstrated that a temporal phase shift (from temporal delay of the AO loop) between phase error (from the atmospheric turbulence) and amplitude error (from the scintillation effect) creates an asymmetry pinned to the wind driven halo in the focal plane image. 

We can define the relative asymmetry factor, $F_{asymmetry}$, as the normalized difference between these two intensities:
\begin{align}
F_{asymmetry} & \, \dot{=} \, \frac{I_+ - I_-}{I_+ + I_-} 
\label{eq-asymmetry_definition}\\
 & = \frac{2 \, sin(2\pi\frac{z}{z_T}) f \Delta r}{\left( sin(2\pi\frac{z}{z_T})\right)^2 + (f \, \Delta r)^2}. 
\label{eq-factor}
\end{align}
This factor is thus between $0$ (no asymmetry) and $1$ (all the light is spread in only one wing). 
\postrefref{\footnote{The asymmetry factor is maximum ($F_{asymmetry} = 1$) when the numerator is equal to $1/2$ (i.~e. $sin(2\pi\frac{z}{z_T}) . f\,\Delta r = 1$: the amplitude error is fully correlated with the phase error), and is minimum ($F_{asymmetry} = 0$) when the numerator is null (null wind speed / no temporal lag: there is no wind driven halo) or equal to infinity (there is no correlation at all between amplitude error and phase error).}}

With current HCI instruments, the WDH has a typical contrast of $10^{-4}$ (see Fig.~\ref{fig-prof}), whereas the scintillation has a typical contrast of $10^{-6}$ \citep{tatarski2016wave} so we can ignore the scintillation term in the denominator \postrefref{and simplify to $sin(\frac{z}{z_T}) \sim \frac{z}{z_T}$}, which yields, after replacing the Talbot length by its expression\postref{, the following approximation}:
\begin{equation}
F_{asymmetry} = \frac{4 z f \lambda}{v_{wind} \, \Delta t} + \mathcal{O} \left( \, \left(\frac{z f \lambda}{v_{wind} \, \Delta t}\right)^2 \, \right).
\label{eq-factorapprox}
\end{equation}
We consequently expect the asymmetry factor to grow linearly with \postref{the spatial frequency, and therefore with} the angular separation to the star. 
%
%
From this demonstration we can already infer a few effects. First, as the interference is taking place between the turbulence residuals and the AO correction lag, any type of coronagraph will reveal the asymmetry of the WDH. 
\postrefref{Second, despite the fact that the Talbot length is $36~\mathrm{km}$ while the jet stream layer is at $12~\mathrm{km}$ altitude, the propagation distance is sufficient enough to convert a small fraction of the phase error into amplitude error and therefore produce the observed asymmetry.}
\postref{Consequently, the higher the altitude of the fast layer, the more asymmetry is produced and, on the contrary, the ground layer does not produce this asymmetry.}
\postrefref{Third, knowing that the amplitude errors are only due to the turbulence whereas the phase delayed error is due to both the wind speed and the AO loop correction speed, the asymmetry varies with temporal parameters as follow: (i) if the AO loop delay $\Delta t$ increases (e.~g. the AO loop is slower) we loose the correlation between the amplitude errors and the delayed phase errors, making the asymmetry smaller; (ii) if the wind speed $v_{wind}$ is higher, all the same the correlation between the amplitude error and the delayed phase error decreases, making the asymmetry smaller. In other word, if the beam shift $\Delta r$ between the turbulent layer and the AO correction increases, the correlation decreases and so does the asymmetry.}
Finally, we expect this asymmetry to increase with wavelength as the Talbot length is inversely proportional to wavelength while the other parameters are independent of wavelength.

\postref{As a consequence, an observation site such as Mauna Kea which suffers from less jet stream compared to observatories located at Paranal in Chile \citep[e.g.][]{sarazin2003} would be beneficial to avoid the wind driven halo\postrefref{\footnote{The Subaru/SCExAO high contrast images do not show the wind driven halo and its asymmetry. This might also be explained by the use of predictive control algorithm based on machine-learning techniques which intends in getting rid of the servolag error \citep{Males2018predcont,Guyon2017predcont}.}} in the high contrast images, and the subsequent asymmetry which may arise depending on the AO correction setting and the speed of the high atmospheric turbulent layers.}

\section{Simulations of the effect}
\label{sec-simu}
In the following, we describe a numerical simulation of an idealized AO system reacting to a simplified atmosphere with a single, high-altitude turbulence layer. The goal is to explore the connection between servolag, scintillation, and the occurrence (or absence) of an asymmetric WDH. The simulations are conducted using the HCIPy package \citep{por2018hcipy}, which is available as open-source software on GitHub\footnote{\url{https://github.com/ehpor/hcipy}}.

We simulated a single atmospheric layer at the altitude of the jet stream, which is then moved across the telescope aperture according to the frozen-flow hypothesis. The light is propagated from the layer to the ground using an angular-spectrum Fresnel propagation code. This light is sensed using a noiseless WFS, which in turn is used to drive a DM. An integral controller with a gain of $0.5$ is assumed. The flattened wavefront is then propagated through a perfect coronagraph \citep{cavarroc2006fundamental} before being focused onto the science camera. We carry out 500 independent short-exposure simulations, whose images are stacked to form the final long-exposure image. A list of the nominal simulation parameters can be found in Table~\ref{tab:simulation_parameters}.

\begin{table}
\caption{Nominal set of parameters used for our simulations.}
\label{tab:simulation_parameters}
\begin{tabular}{ll}
\hline
\textbf{Parameter name} & \textbf{Value} \\ \hline
Wavelength & $2.2~\mathrm{\mu m}$ (K-band) \\
Pupil diameter & $8~\mathrm{m}$ \\
Seeing & $r_0=20~\mathrm{cm}$ at $500~\mathrm{nm}$ \\
Outer scale & $22~\mathrm{m}$ \\
Jet stream height & $12~\mathrm{km}$ \\
Jet stream velocity & $30~\mathrm{m/s}$ \\
AO system loop speed & $1380~\mathrm{Hz}$ \\
AO system controller & Integral control \\
AO system loop gain & $0.5$ for all modes \\
Corrected modes & 1000 modes \\
Number of actuators & $40\times40$ rectangular grid \\
Influence functions & Gaussian with $\sigma=22~\mathrm{cm}$ projected \\
Coronagraph & Perfect \citep{cavarroc2006fundamental} \\
Wavefront sensor & Noiseless \\
\hline
\end{tabular}
\end{table}

\begin{figure}
\centering
\includegraphics[width=\columnwidth]{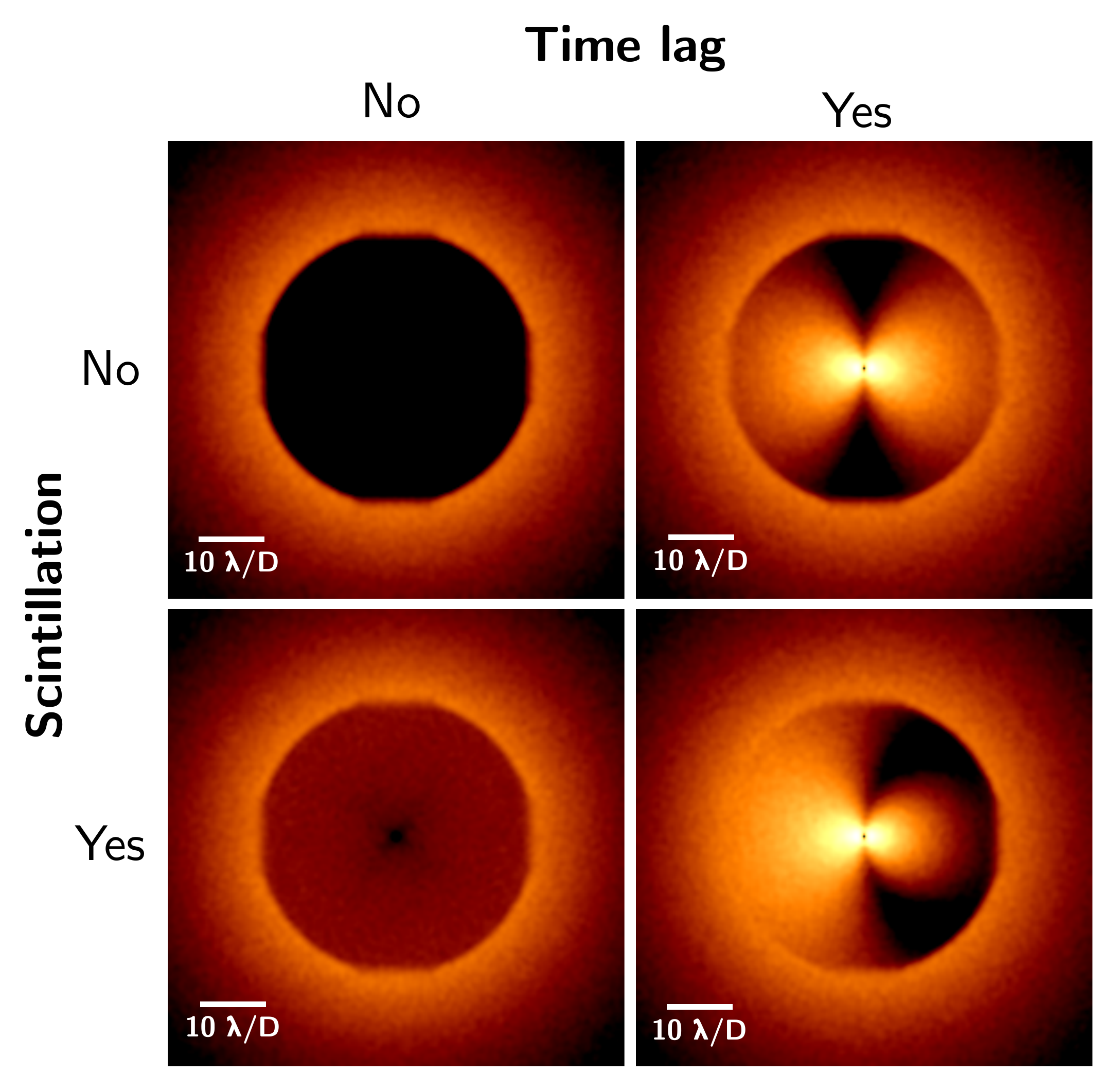}
\caption{Simulated images using HCIPy for the parameters gathered on Tab.~\ref{tab:simulation_parameters}. The images with no time lag were produced with an infinite AO loop speed. Only a time-lagged WDH and scintillation yields an asymmetric coronagraphic PSF. The images are purposely stretched in intensity to highlight the asymmetry and scintillation (log scale).}
\label{fig:asymmetry_matrix}
\end{figure}

Fig.~\ref{fig:asymmetry_matrix} shows the coronagraphic simulated images obtained with or without AO lag and with or without scintillation. As expected, only the combination of both amplitude error and AO servolag error leads to an asymmetric WDH. 

\begin{figure}
\resizebox{\hsize}{!}{\includegraphics{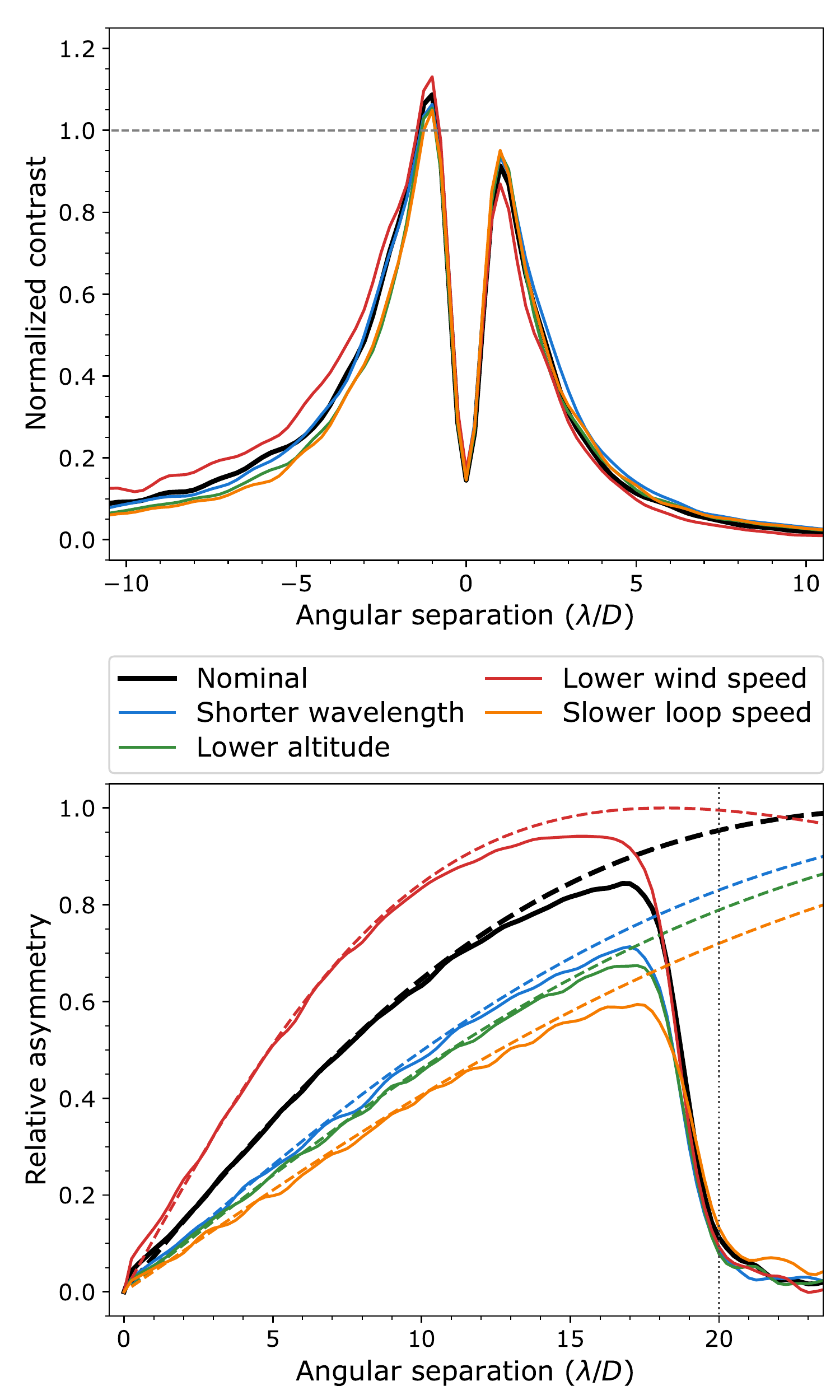}}
\caption{\postref{Radial profiles along the wind direction (top) and asymmetry factor as defined at Eq.~(\ref{eq-asymmetry_definition}) (bottom) for the simulated data sets. Solid black line is using the nominal parameters of Tab.~\ref{tab:simulation_parameters} (image shown in the bottom right of Fig.~\ref{fig:asymmetry_matrix}). Other lines differ in one parameter playing a role in the asymmetry: shorter wavelength (H-band, $\lambda=1.6~\mathrm{\mu m}$, blue line), lower altitude ($z=8~\mathrm{km}$, green line), lower wind speed ($v_{wind}=20~\mathrm{m/s}$, red line) and slower AO loop frequency ($f_{AO}=800~\mathrm{Hz}$, yellow line). The DM cutoff frequency is at $20~\mathrm{\lambda/D}$ (dotted gray line). In the bottom plot, solid lines indicate the asymmetry from the simulated images, and dashed lines show the prediction from Eq.~(\ref{eq-factorapprox}).}}
\label{fig-profsimu}
\end{figure}

Fig.~\ref{fig-profsimu} shows \postref{the radial profile of the simulated images along the wind direction (top) and the corresponding asymmetry factor as defined at Eq.~(\ref{eq-asymmetry_definition}) (bottom)}, as a function of the separation to the star, where we indeed observe that the asymmetry grows linearly with the separation. 
We also demonstrate that indeed the scintillation from the jet stream layer at $12~\mathrm{km}$ altitude is enough to create the asymmetry of the wind driven halo and that lower altitude layers create less asymmetry. 
\postref{As expected from the approximation of Eq.~(\ref{eq-factorapprox}),} our simulations also show that the asymmetry is stronger when the wind speed decreases or when the AO loop frequency \postref{decreases (for a fixed AO loop gain)}.
We also checked that the asymmetry factor is indeed higher at longer wavelengths.

In a forthcoming paper, we will compare this analysis to on-sky images obtained with SPHERE, which involves isolating the contribution of the WDH in the image since other error terms are hiding these trends. 

\section{Conclusions}
In this letter we pointed out the presence of an asymmetry of the wind driven halo that is revealed in high contrast images. 
We described and demonstrated its origin to being due to interference between AO correction lag (\postrefref{delayed phase} error) and scintillation (amplitude errors). We supported our demonstration by simulating this effect using an end-to-end simulator. 
From those simulations we confirmed the expected behavior of the asymmetry with different atmospheric turbulence conditions, XAO correction, and imaging wavelength. \postrefref{We further demonstrated, that the jet stream layer is the main culprit for this aberration since it is responsible for both servolag error (being a fast layer) and scintillation (being a high altitude layer). Therefore, an observing site with weak or no jet stream would get around this aberration.}

While the current letter focuses on exploring the origin of the wind driven halo asymmetry \postrefref{so as to better understand our current observations and AO systems for future designs}, a more quantitative analysis of its implication on high contrast imaging capabilities and potential mitigation strategies will be detailed in a separate paper. \postrefref{Indeed, the servolag error, when present, is now one of the major effect limiting the high contrast capabilities of the current instruments (along with the low wind effect, the non-common path aberrations and residual tip-tilt errors). Knowing that this wind driven halo shows an asymmetry makes it more difficult to deal with in post-processing (as using for instance the residual phase structure functions yields a symmetric phase error or that most filters have a symmetric effect).}

Now that this effect is acknowledged and demonstrated, next step is to take it into account within end-to-end XAO simulators \cite[e.g. COMPASS or SOAPY,][]{gratadour2014compass,reeves2016soapy}
or analytical simulators \citep[e.g. PAOLA,][]{Jolissaint2010PAOLA} and more generally in XAO error budgets, when used in the HCI framework. 
This study gives insights into the instrument operations, essential to designing optimal post-processing techniques or AO predictive control tools that both aim to get rid of the servolag error signature \citep[e.g.][]{Males2018predcont, Correia2018predcont}. This effect is also important to design the next generation of high contrast instruments \citep[e.g. MagAOX][or giant segmented mirror telescopes instruments dedicated to HCI]{Close2018MagAOX} or to lead the upgrades of existing high contrast instruments \citep[e.g. GPI or SPHERE, ][]{Chilcote2018GPIupgrade, Beuzit2018sphereupgrade} \postrefref{by, for instance, adding a second DM to correct for the scintillation.}

\begin{acknowledgements}
We would like to thank the anonymous referee for his/her comments which helped clarifying the letter. The authors also would like to thank J.~Lozi and O.~Guyon for the discussions about the wind driven halo in the Subaru/SCExAO images. EHP acknowledges funding by The Netherlands Organisation for Scientific Research (NWO) and the S\~{a}o Paulo Research Foundation (FAPESP). NAB acknowledges UK Research \& Innovation Science and Technologies Facilities Council funding (ST/P000541/1).
\end{acknowledgements}


\bibliographystyle{aa} 
\bibliography{my_biblio_ah}

\end{document}